\documentclass[letterpaper, 10 pt, conference]{ieeeconf}

\usepackage{amsmath,amssymb,epsfig,multicol}
\usepackage{epsf}

\IEEEoverridecommandlockouts

\begin{document}

\title{\LARGE \bf Reliability Bounds for Delay-Constrained Multi-hop Networks} 

\author{\"Ozg\"ur Oyman, ~\IEEEmembership{Member,~IEEE} 
\thanks{The author is with the Corporate Technology Group,
Intel Corporation, Santa Clara, CA 95054, U.S.A. (email: ozgur.oyman@intel.com)}
}


%


\maketitle

\begin{abstract}
We consider a linear multi-hop network composed of multi-state discrete-time memoryless channels over each hop, with orthogonal time-sharing across hops under a half-duplex relaying protocol. We analyze the probability of error and associated reliability function \cite{Gallager68} over the multi-hop network; with emphasis on random coding and sphere packing bounds, under the assumption of point-to-point coding over each hop. In particular, we define the system reliability function for the multi-hop network and derive lower and upper bounds on this function to specify the reliability-optimal operating conditions of the network under an end-to-end constraint on the total number of channel uses. Moreover, we apply the reliability analysis to bound the expected end-to-end latency of multi-hop communication under the support of an automatic repeat request (ARQ) protocol. Considering an additive white Gaussian noise (AWGN) channel model over each hop, we evaluate and compare these bounds to draw insights on the role of multi-hopping toward enhancing the end-to-end rate-reliability-delay tradeoff.
\end{abstract}

\section{Discrete-Time Memoryless Multi-hop Network}

\begin{figure} [t]
\begin{center}
\includegraphics[width=3.3in, keepaspectratio]{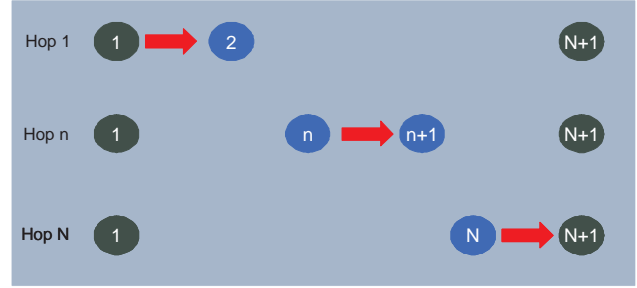}
\end{center}
\caption{Linear half-duplex multihop network model.}
\label{linear_net1}
\end{figure}

We define a linear multi-hop network as a network where a pair of source and destination terminals communicate with each other by routing their data through multiple intermediate relay terminals. We assume that the linear multi-hop network consists of $N+1$ terminals; the source terminal is identified as ${\cal T}_1$, the destination terminal is identified as ${\cal T}_{N+1}$, and the intermediate terminals are identified as ${\cal T}_2$-${\cal T}_{N}$, where $N$ is the number of hops along the transmission path. As terminals can often not transmit and receive at the same time, we only focus on time-division based (half duplex) relaying. In particular, we consider a simple $N$-hop decode-and-forward protocol, where, at hop $n$, relay terminal ${\cal T}_{n+1}, \, n=1,...,N-1$ hears and attempts to fully decode the data signal transmitted from terminal ${\cal T}_{n}$ and forwards its re-encoded version over hop $n+1$ to terminal ${\cal T}_{n+2}$, as depicted in Fig. \ref{linear_net1}.

To model block-coded communication over the linear multi-hop network, a $((e^{Q_1R_1},...,e^{Q_NR_N}),\{Q_n\}_{n=1}^N,Q)$ code ${\cal C}_Q$ is defined by a codebook of
$\sum_{n=1}^N e^{Q_nR_n}$ codewords such that $R_n$ is the rate of communication over hop
$n$ (in nats per channel use), $Q_n$ is the coding blocklength over hop $n$ and $Q = \sum_{n=1}^N Q_n$ is the fixed total number of channel uses over the multi-hop link, representing a delay-constraint in the end-to-end sense, i.e., the time-division based $N$-hop communication
takes place over the total duration of $\sum_{n=1}^N Q_n = Q$ symbol periods. Let ${\cal S}_{Q_n}$ to be the set of all sequences of length $Q_n$ that can be transmitted on the channel over hop $n$ and ${\cal Y}_{Q_n}$ to be the set of all sequences of length $Q_n$ that can be received. The codebook for multi-hop transmissions is determined by the encoding functions 
$\phi_n,\,n=1,...,N$, that map each message $w_n\in {\cal W}_n = \{1,...,e^{Q_nR_n}\}$ over hop $n$ 
to a transmit codeword
${\bf s}_n = \left[ \,s_{n,1},\, . . .\, , s_{n,Q_n}\,
\right] \in {\Bbb C}^{1 \times Q_n}$, where $s_{n,q} \in {\cal S}_{1}$ is the transmitted
symbol over hop $n$ at time $\sum_{m=1}^{n-1}Q_m+q$, which is drawn according to a continuous distribution $p(s)$ defined on a complex-valued infinite alphabet. The strictly positive time-sharing constant $\lambda_n = Q_n/Q > 0$ is defined as the fractional time over which the transmission and reception over hop $n$ is active. 

Let $P({\bf y}_n | {\bf s}_n, {\theta}_n)$, for ${\bf s}_n \in {\cal S}_{Q_n}$ and ${\bf y}_n \in {\cal Y}_{Q_n}$, be the conditional probability density of receiving sequence ${\bf y}_n$, given that ${\bf s}_n$ was transmitted and the channel state over hop $n$ is given by ${\theta}_n$, which is drawn from an arbitrary-size state space ${\Theta}$ of channel states. Thus, our multi-state channel model concentrates on the quasi-static regime, in which, once drawn, the channel states $\{\theta_n\}_{n=1}^N$ remain fixed for the entire duration of the respective hop transmissions over blocklengths $\{Q_n\}_{n=1}^N$.
The channel over each hop $n=1,...,N$ is assumed to be a discrete-time infinite-alphabet memoryless channel (DTMC) satisfying
$$
P({\bf y}_n | {\bf s}_n, \theta_n) = \prod_{q=1}^{Q_n} p(y_{n,q} | s_{n,q}, \theta_{n}).
$$
for all ${\bf s}_n \in {\cal S}_{Q_n}$, and ${\bf y}_n \in {\cal Y}_{Q_n}$ and for all $Q_n$, where $p(y|s,\theta)$ is the conditional probability density of receiving output symbol $y$ chosen from support set ${\cal Y}_1$ given that the input symbol was $s$ chosen from support set ${\cal S}_1$ and the channel state was $\theta$ chosen from support set ${\Theta}$. The channels over different hops are assumed to exhibit independent and identical (i.e., i.i.d. across $n$) statistical behavior. We shall assume that each transmitting terminal ${\cal T}_{n},\,n=1,...,N$ and each receiving terminal ${\cal T}_{n+1},\,n=1,...,N$ possesses the knowledge of the channel state $\theta_n$ over hop $n$, for $n=1,...,N$, which allows transmission rate $R_n$ to be chosen in a way that guarantees a desired level of reliability for a given coding blocklength $Q_n$, e.g., $R_n$ could be chosen to enable perfectly reliable communication provided that coding blocklength $Q_n$ is large. It should be emphasized that we only assume the presence of local channel state information (CSI) at the terminals, where each terminal knows perfectly the transmit and receive CSI regarding its neighboring links only, and our results do not require the presence of global CSI at the terminals over the multi-hop network. 

Each receiving terminal employs a decoding function $\psi_n \, , \,n=1,...,N$ to perform the mapping ${\Bbb C}^{1
\times Q_n} \rightarrow \hat{w}_n \in {\cal W}_n$ based on its observed
signal vector ${\bf y}_n = \left[ \, y_{n,1},\, . . .\, , y_{n,Q_n}\, \right]$ according to the maximum likelihood decoding rule, where $y_{n,q} \in {\cal Y}_{1}$ is the received symbol over hop $n$ at time $\sum_{m=1}^{n-1}Q_m+q$.
The codeword error probability for the $n$-th hop is given by $\epsilon_n = {\Bbb
P}(\psi_n({\bf y}_n) \neq w_n)$. An $N$-tuple of multi-hop rates $(R_1,...,R_N)$ is
achievable if there exists a sequence of $((e^{Q_1R_1},...,e^{Q_NR_N}), \{Q_n\}_{n=1}^N, Q)$
codes $\{{\cal C}_Q:Q=1,2,...\}$ with $Q=\sum_{n=1}^N Q_n$, $\lambda_n >0, \forall n$, and vanishing $\epsilon_n,\, \forall n$.
 
The ensemble of codes $\{{\cal S}_{Q_n}\}_{n=1}^N$ to be transmitted over the multi-hop link is chosen by defining a probability measure $P({\bf s}_n)$ on the set ${\cal S}_ {Q_n}$ of possible input sequences to the channel over hop $n$. An ensemble of $M_n = e^{Q_nR_n}$ codewords is constructed for transmission over hop $n$ by picking each codeword independently, according to the probability measure $P({\bf s}_n)$. Thus the probability associated with a code consisting of codewords ${\bf s}_1$,...,${\bf s}_{M_n}$ is $\prod_{m=1}^{M_n} P({\bf s}_{m})$. Furthermore, we restrict the class of ensembles of codes under consideration to those in which each symbol of each codeword is chosen independently of all other symbols with a probability measure $p(s)$, i.e. $P({\bf s}_n) = \prod_{q=1}^{Q_n} p(s_{n,q})$.

\section{Reliability Bounds for Multi-hop Networks}

Capacity and mutual information in linear multi-hop networks have been characterized in \cite{Reznik04}-\nocite{Tuninetti2005}\nocite{Laneman05}\cite{Oyman06b}. In particular, based on the model introduced by Section I, we characterized in \cite{Oyman06b} the end-to-end mutual information $I$ (in nats per channel use) of the linear multi-hop network conditional on the channel states $\{\theta_n\}_{n=1}^N$ as 
\begin{eqnarray}
I(\{\theta_n\}_{n=1}^N) &=& \frac{1}{Q} \, \max_{\sum_{n=1}^N Q_n = Q} \, \min_n \,\, \sum_{q=1}^{Q_n} I(S_{n,q};Y_{n,q}\,|\,\theta_n) \nonumber \\
&=& \max_{\sum_{n=1}^N \lambda_n = 1} \min_n \left\{ \lambda_n 
I_n(S_n;Y_n\,|\,\theta_n) \right\},
\label{cap_minimax}
\end{eqnarray}
where $\lambda_{n} \in [0,1]$ is the fractional time of the channel corresponding to hop $n$, i.e. $\lambda_n=Q_n/Q$, $I_n(S_{n,q};Y_{n,q}\,|\,\theta_n)$ is the conditional mutual information (a random variable that depends on the channel state $\theta_n$) over hop $n$ at time $\sum_{m=1}^{n-1} Q_m + q$ and the second line of (\ref{cap_minimax}) follows from the fact that the transmit symbols $\{s_{n,q}\}$ are drawn in an independent and identically distributed (i.i.d.) fashion across $q=1,...,Q_n$ based on the probability density $p(s)$ and that for any given state $\theta_n$, the channel over hop $n$ that maps the input symbol vector ${\bf s}_n$ to the output symbol vector ${\bf y}_n$ acts i.i.d. across $q=1,...,Q_n$ (based on the DTMC model) dictated by the conditional probability density function $p(y|s,\theta_n)$. This formulation assumes that the channel state $\theta_n$ is known by both transmit terminal ${\cal T}_n$ and receive terminal ${\cal T}_{n+1}$. The minimax point of (\ref{cap_minimax}) is achieved by choosing
$(\lambda_1^*,...,\lambda_N^*)$ such that
$$
\lambda_n^* = \frac{\prod_{k \neq n} I_k(S_k;Y_k\,|\,\theta_k)}{\sum_{k=1}^N \prod_{k \neq n} 
I_k(S_k;Y_k\,|\,\theta_k)},\,\,\,\,\,\,\,n=1,...,N, 
$$
which results in
\begin{eqnarray}
I(\{\theta_n\}_{n=1}^N) &=& \frac{\prod_{n=1}^N I_n(S_n;Y_n\,|\,\theta_n)}{\sum_{k=1}^N
\prod_{k \neq n} I_k(S_k;Y_k\,|\,\theta_k)} \nonumber \\
&=&\frac{1}{\sum_{n=1}^N \frac{1}{I_n(S_n;Y_n\,|\,\theta_n)}}.
\label{multihop_cap}
\end{eqnarray}

Knowing the capacity of a channel is not always sufficient. One may be interested in how hard it is to get close to this capacity while guaranteeing a certain level of reliability. In particular, one of the most important tradeoffs in the design of reliable adhoc networks and multi-hop communication systems is the tradeoff between end-to-end system delay and probability of error for a given set of achievable rates over multiple hops. Upper and lower bounds on the probability of decoding error based on error exponents and sphere packing exponents provide a partial answer to this question by giving reliability bounds achievable by the best block codes of a certain length and rate, which help to quantify the exponential rate of decrease in error probability with increasing blocklength. Our objective here is to analyze these reliability exponent functions over the discrete-time memoryless multi-hop network, and provide insights on the tradeoffs among achievable rates, error probability and end-to-end delay in multi-hop networks.

Assuming point-to-point coding over each hop, we define the reliability function of the DTMC over hop $n$, $E_n(R_n,\theta_n)$, as \cite{Gallager68}
$$
E_n(R_n,\theta_n) = \lim_{Q_n \rightarrow \infty} - \frac{\ln \left(P_{e,n}(Q_n,R_n,\theta_n)\right)}{Q_n}\,\,\,\,\,n=1,...,N
$$
where $P_{e,n}(Q_n,R_n,\theta_n)$ is the minimum codeword error probability over the $n^{\mathrm{th}}$ hop for all codes of blocklength $Q_n$ and transmission rate $R_n$ (in nats per channel use) given the channel state $\theta_n$. For each DTMC in the linear multi-hop network, the minimum probability of decoding error $P_{e,n}$ for codes of blocklength $Q_n$ can be bounded for any rate below the capacity between the limits
$$
e^{-Q_n\left[E_{sp,n}(R_n,\theta_n)+O(Q_n)\right]} \leq P_{e,n} \leq e^{-Q_nE_{r,n}(R_n,\theta_n)}.
$$
In this expression, $E_{r,n}(R_n,\theta_n)$ and $E_{sp,n}(R_n,\theta_n)$ are lower and upper bounds, respectively, on the reliability function $E_n(R_n,\theta_n)$, and are known as the error exponent and sphere packing exponent, respectively and $O(Q_n)$ is a function going to $0$ with increasing $Q_n$. It should also be noted that for any given $Q_n$, $E_{r,n}(R_n,\theta_n) = E_{sp,n}(R_n,\theta_n) = E_n(R_n,\theta_n)$ for $R_{cr,n} \leq R_n \leq I_n(\theta_n)$ where $R_{cr,n}$ is the critical rate for hop $n$ and $I_n(\theta_n)$ is the maximum achievable mutual information conditional on the channel state $\theta_n$ (or Shannon capacity, which exists since the channel state $\theta_n$ is known by both transmit terminal ${\cal T}_n$ and receive terminal ${\cal T}_{n+1}$) over hop $n$ expressed by (\ref{mutu_theta}),
\begin{figure*}[!ht]
\begin{equation}
I_n(\theta_n) = \underset{p(s)}{\sup} \int_{{\cal S}_1} \int_{{\cal Y}_1} p(s) p(y|s,\theta_n) \ln \left(\frac{p(y|s,\theta_n)}{\int_{{\cal S}_1} p(s')p(y|s',\theta_n)ds'}\right) dy \, ds
\label{mutu_theta}
\end{equation}
\hrule
\end{figure*}
and thus the exponential dependence of error probability on blocklength quantified by the reliability function $E_n(R_n,\theta_n)$ is known exactly in this range (i.e., for all rates above the critical rate). The random coding error exponent $E_{r,n}(R_n,\theta_n)$ over hop $n$ is given by
\begin{equation}
E_{r,n}(R_n,\theta_n) = \underset{0 \leq \rho \leq 1}{\max} \,\left[ -\rho R_n + E_{0,n}(\rho,\theta_n) \right],
\label{err_exp}
\end{equation}
where in turn $E_{0,n}(\rho,\theta_n)$ is given by the supremum over all input distributions $p(s)$ satisfying (\ref{integ_err}).
\begin{figure*}[!ht]
\begin{equation}
E_{0,n}(\rho,\theta_n) = \underset{p(s)}{\sup} \,\left\{ -\ln \int_{{\cal Y}_{1}} \left[ \int_{{\cal S}_{1}} p(s) p(y|s,\theta_n)^{1/(1+\rho)}ds \right]^{1+\rho} dy \right\}. 
\label{integ_err}
\end{equation}
\hrule
\end{figure*}
The sphere packing exponent $E_{sp,n}(R_n,\theta_n)$ controlling the lower bound to error probability is given by
\begin{equation}
E_{sp,n}(R_n,\theta_n) = \underset{0 < \rho < \infty}{\max} \,\left[ -\rho R_n + E_{0,n}(\rho,\theta_n) \right].
\label{sp_exp}
\end{equation}
We see that the only difference between the error exponent and sphere packing exponent is in the range for which the optimization over $\rho$ is performed. If the maximizing $\rho$ lies between $0$ and $1$, then $E_{r,n}(R_n,\theta_n) = E_{sp,n}(R_n,\theta_n)=E_n(R_n,\theta_n)$ and therefore the upper and lower bounds to error probability agree in their exponential dependence on $Q_n$. Thus, if $E_{r,n}(R_n,\theta_n)=E_{sp,n}(R_n,\theta_n)$ for one value of $R_n$, then equality also holds for all larger values of $R_n$. We define the critical rate over hop $n$, denoted by $R_{cr,n}$, as the smallest $R_n$ such that $E_{sp,n}(R_n,\theta_n) = E_{r,n}(R_n,\theta_n)$, which thus holds for $R_{cr,n} \leq R_n \leq I_n(\theta_n)$. It should be observed that for all $R_n < R_{cr,n}$, the maximum over $\rho$ of $E_{r,n}(R_n,\theta_n)$ occurs at $\rho=1$. Setting the partial derivative of the bracketed part of (\ref{err_exp}) equal to $0$, we get
\begin{equation}
R_n = \frac{\partial E_{0,n}(\rho_n,\theta_n)}{\partial \rho}.
\label{rho_deriv}
\end{equation}
If some $\rho_n$ in the range $0 \leq \rho_n \leq 1$ satisfies (\ref{rho_deriv}), then that $\rho_n$ must maximize (\ref{err_exp}). Furthermore, since $\partial E_{0,n}(\rho,\theta_n)/\partial \rho$ is nonincreasing with $\rho$, a solution to (\ref{rho_deriv}) over the interval $[0,1]$ exists if $R_n$ lies in the range
$$
\left. \frac{\partial E_{0,n}(\rho,\theta_n)}{\partial \rho} \right|_{\rho=1} \leq R_n \leq I_n(\theta_n).
$$
In this range it is most convenient to use this equation to relate $E_{r,n}(R_n,\theta_n)$ and $R_n$ parametrically as functions of $\rho_n$. This gives us
\begin{equation}
E_{r,n}(R_n,\theta_n) = E_{0,n}(\rho_n,\theta_n) - \rho_n \frac{\partial E_{0,n}(\rho_n,\theta_n)}{\partial \rho}
\label{err1}
\end{equation}
and
\begin{equation}
R_n = \frac{\partial E_{0,n}(\rho_n,\theta_n)}{\partial \rho}, \,\,\,0 \leq \rho_n \leq 1.
\label{err2}
\end{equation}
We therefore see that $R_n$ is a strictly decreasing function of $\rho_n$. For $R_n < \left. \partial E_{0,n}(\rho,\theta_n) / \partial \rho \right|_{\rho = 1}$, the parametric equations above are not valid. In this case, the function $-\rho R_n + E_{0,n}(\rho,\theta_n)$ increases with $\rho$ in the range $0 \leq \rho \leq 1$, and therefore the maximum occurs at $\rho = 1$. Thus
\begin{equation}
E_r(R_n,\theta_n) = E_{0,n}(1,\theta_n) - R_n \,\,\,\,\mathrm{for}\,\,\,\,R_n<\left.\frac{\partial E_{0,n}(\rho,\theta_n)}{\partial \rho}\right|_{\rho=1}.
\label{err3}
\end{equation}
For the sphere packing exponent in (\ref{sp_exp}), since the maximizing $\rho$ can be any positive number, we have
\begin{equation}
E_{sp,n}(R_n,\theta_n) = E_{0,n}(\rho_n,\theta_n) - \rho_n \frac{\partial E_{0,n}(\rho_n,\theta_n)}{\partial \rho}
\label{sp1}
\end{equation}
and
\begin{equation}
R_n = \frac{\partial E_{0,n}(\rho_n,\theta_n)}{\partial \rho}.
\label{sp2}
\end{equation}

{\bf System Reliability Function for Multi-hop Network.} What remains is to use the per-hop reliability function and associated lower and upper bounds through error exponents and sphere packing exponents to develop a theoretical framework for the notion of a system-wide reliability concept that accounts for error events over all hops over the linear network. To this end, we define the system reliability function as a measure for the probability of error in an end-to-end sense over the multi-hop network by summing the error probabilities $\{P_{e,n}\}_{n=1}^N$ over each hop. A union bound interpretation can also be attached to this measure, as the end-to-end error probability can be upperbounded by the sum of individual link error probabilities. Thus the system probability of error for the linear multi-hop network is defined as
$$
P_e \left(Q,\{R_n\}_{n=1}^N,\{\theta_n\}_{n=1}^N \right) = \sum_{n=1}^N P_{e,n}(Q_n,R_n,\theta_n),
$$
and consequently, the end-to-end system reliability function $E_{sys}\left(\{R_n\}_{n=1}^N,\{\theta_n\}_{n=1}^N\right)$ is defined as given in (\ref{sys_rel}).
\begin{figure*}[!ht]
\begin{equation}
E_{sys}\left(\{R_n\}_{n=1}^N,\{\theta_n\}_{n=1}^N \right) = \lim_{Q \rightarrow \infty} \, - \frac{\ln \left(P_{e}(Q,\{R_n\}_{n=1}^N,\{\theta_n\}_{n=1}^N)\right)}{Q} \\
\label{sys_rel}
\end{equation}
\hrule
\end{figure*}
Based on the random coding upper bound and sphere packing lower bound, the system probability of error can be bounded by the expression in (\ref{sandwich}),
\begin{figure*}[!ht]
\begin{equation}
\sum_{n=1}^N \exp \left( - Q_n E_{sp,n}(R_n,\theta_n) \right) \leq P_e\left(Q,\{R_n\}_{n=1}^N,\{\theta_n\}_{n=1}^N \right) \leq \sum_{n=1}^N \exp \left( - Q_n E_{r,n}(R_n,\theta_n) \right)
\label{sandwich}
\end{equation}
\hrule
\end{figure*}
and consequently, the system reliability function can be written as in (\ref{sys_bnd}).
\begin{figure*}[!ht]
\begin{equation}
\lim_{Q \rightarrow \infty} - \frac{ \ln \left( \sum_{n=1}^N \exp \left( - Q_n E_{sp,n}(R_n,\theta_n) \right) \right)}{Q} \leq  E_{sys}\left(\{R_n\}_{n=1}^N,\{\theta_n\}_{n=1}^N\right) \leq \lim_{Q \rightarrow \infty} - \frac{ \ln \left( \sum_{n=1}^N \exp \left( - Q_n E_{r,n}(R_n,\theta_n) \right) \right)}{Q}
\label{sys_bnd}
\end{equation}
\hrule
\end{figure*}

{\bf Reliability-Optimal Block Allocation.} In this setting, our objective is to choose coding blocklengths $\{Q_n\}_{n=1}^N$ optimally in order to maximize multi-hop link reliability. To this end, we should now minimize the upper and lower bounds on $P_e$ subject to the constraint  $\sum_{n=1}^N Q_n = Q$, which translates into the maximization of the system reliability function leading to (\ref{sys_rel2}). 
\begin{figure*}[!ht]
\begin{equation}
E_{sys}\left(\{R_n\}_{n=1}^N,\{\theta_n\}_{n=1}^N \right) = \lim_{Q \rightarrow \infty} \,\,\underset{\sum_{n=1}^N Q_n = Q} {\max} - \frac{\ln \left(P_{e}(Q,\{R_n\}_{n=1}^N,\{\theta_n\}_{n=1}^N)\right)}{Q} \\
\label{sys_rel2}
\end{equation}
\hrule
\end{figure*}
Carrying out the optimization using Lagrange multipliers and utilizing from the error exponent and sphere packing exponent lower and upper bounds, respectively, on the per-hop reliability function, we find that the optimal blocklengths $\{Q_n^*\}_{n=1}^N$ can be bounded as in (\ref{opt_blk}),
\begin{figure*}[!ht]
\begin{equation}
\min \left\{ \frac{ \ln E_{r,n}(R_n,\theta_n) - \lambda_r}{E_{r,n}(R_n,\theta_n)}, \frac{ \ln E_{sp,n}(R_n,\theta_n) - \lambda_{sp}}{E_{sp,n}(R_n,\theta_n)} \right\} < Q_n^{*} < \max \left\{ \frac{ \ln E_{r,n}(R_n,\theta_n) - \lambda_r}{E_{r,n}(R_n,\theta_n)}, \frac{ \ln E_{sp,n}(R_n,\theta_n) - \lambda_{sp}}{E_{sp,n}(R_n,\theta_n)} \right\} 
\label{opt_blk}
\end{equation}
\hrule
\end{figure*}
where constants $\lambda_r$ and $\lambda_{sp}$ are given by
$$
\lambda_{r} = \left( \sum_{n=1}^N \frac{1}{E_{r,n}(R_n,\theta_n)}\right)^{-1} \, \left( \sum_{n=1}^N \frac{\ln E_{r,n}(R_n,\theta_n)}{E_{r,n}(R_n,\theta_n)} - Q \right),
$$
$$
\lambda_{sp} = \left( \sum_{n=1}^N \frac{1}{E_{sp,n}(R_n,\theta_n)}\right)^{-1} \, \left( \sum_{n=1}^N \frac{\ln E_{sp,n}(R_n,\theta_n)}{E_{sp,n}(R_n,\theta_n)} - Q \right).
$$
We observe from this solution that under the end-to-end delay constraint, the reliability-optimal solution favors {\it error balancing across multiple hops}, by allocating blocks across different links to ensure the same exponential decay of the individual link error probabilities.

{\bf Information-Continuous Block Allocation.} Alternatively, we can consider the {\it information balancing} allocation of blocks across multiple hops, by letting $M_n=e^{Q_nR_n}=M$ and therefore fixing the number of transmitted codewords over all hops at a constant value $M$, which ensures information continuity and no data accumulation at any relay terminal. Thus, the code blocklengths in this approach equal $Q_n = \lfloor \frac{\ln M}{R_n} \rfloor$. Under the constraint $\sum_{n=1}^N Q_n = Q$, this implies that the number of codewords needs to be chosen as 
\begin{equation}
M = \left\lfloor \exp \left(\frac{Q}{\sum_{n=1}^N \frac{1}{R_n}} \right) \right\rfloor.
\label{num_words}
\end{equation} 
It should be noted that the information-continuous block allocation also leads to the optimal time-sharing solution that achieves multi-hop network capacity in (\ref{multihop_cap}). This is because the imposed condition $\lambda_nI_n(S_n;Y_n\,|\,\theta_n)=\lambda_kI_k(S_k;Y_k\,|\,\theta_k), \,1 \leq n < k \leq N$ in (\ref{cap_minimax}) implies that for the optimal time-sharing solution under rate-adaptation (i.e., $R_n=I_n(S_n;Y_n\,|\,\theta_n),\,\forall n$), it also holds that $Q_nR_n=Q_kR_k, \,1 \leq n < k \leq N$ and the time-sharing coefficient can be determined as $\lambda_n = Q_n / Q = R/R_n$. Consequently, the capacity-optimal rate-adaptive relaying technique that achieves (\ref{multihop_cap}) arranges the multi-hop transmissions such that the hops with poor channel conditions transmit relatively longer packets than the hops experiencing good channel conditions. 

{\bf Distributed Implementation.} To implement the information-balancing solution over time-varying random channels (e.g., fading wireless channels), where the maximum achievable mutual information $\{I_n(\theta_n)\}_{n=1}^N$ over each hop of the linear network becomes a random variable, the transmit terminal ${\cal T}_n$ over hop $n$ only needs to know the values of the fixed number of codewords $M$ and channel state $\theta_n$ in order to design its transmit codebook since from these parameters, data rate $R_n$ and coding blocklength $Q_n$ can be determined. The knowledge of global channel state information (CSI) (i.e. CSI for all DTMCs in the multi-hop network given by $\{\theta_n\}_{n=1}^N$) is not required at every terminal, which implies significantly reduced messaging overhead. The information on $\theta_n$ over hop $n$ can be obtained by ${\cal T}_n$ through CSI feedback from the neighboring terminal ${\cal T}_{n+1},\,n=1,...,N$. On the other hand, the parameter $M$ depends on the channel conditions over all links, which may be computed in a distributed fashion prior to data transmission using a routing algorithm (e.g., destination-sequenced
distance-vector (DSDV) algorithm \cite{Perkins94}) where the cost of the link over hop $n$ is represented by the metric $1/R_n$, which is also known as the {\it expected transmission time (ETT)} \cite{Draves04} in the networking literature. Such a distributed approach involves the end-to-end propagation of a single parameter; only requiring neighbor-to-neighbor message passing of the accumulated multi-hop link cost metric which is updated by each terminal with the addition of the cost of the last hop. Once the total route cost $\sum_{n=1}^N \frac{1}{R_n}$ has been determined by one of the end terminals, the value of $M$ can be computed and broadcasted to all the terminals in the linear multi-hop network.

Similar distributed approaches can be applied to perform reliability-optimal block allocation over multi-hop networks so that, for instance, terminal ${\cal T}_n$ can design its transmit codebook over hop $n$ using only the local channel state $\theta_n$ and relation in (\ref{opt_blk}), which would be sufficient to choose $Q_n$ and $R_n$ to minimize the system probability of error. To enable such reduced overhead computation without global CSI, only the knowledge of $\lambda_r$ and $\lambda_{sp}$ should be present at all terminals before the beginning of transmissions over the multi-hop network; which are functions of all channel states $\{\theta_n\}_{n=1}^N$. Again, the DSDV algorithm can be executed to compute relevant end-to-end link cost metrics to obtain these parameters by neighbor-to-neighbor message passing, in which case the per-hop link cost metrics are now defined as functions of the error exponents and sphere packing exponents over the corresponding hops, i.e., relevant metrics of interest would be $1/E_{r,n}(R_n,\theta_n)$, $1/E_{sp,n}(R_n,\theta_n)$, $\ln(E_{r,n}(R_n,\theta_n))/E_{r,n}(R_n,\theta_n)$ and $\ln(E_{sp,n}(R_n,\theta_n))/E_{sp,n}(R_n,\theta_n)$. 

In Section IV, we will investigate the impact of different methods for per-hop blocklength selection on the end-to-end reliability of multi-hop communication. 

\begin{figure} [t]
\begin{center}
\includegraphics[width=3.3in, keepaspectratio]{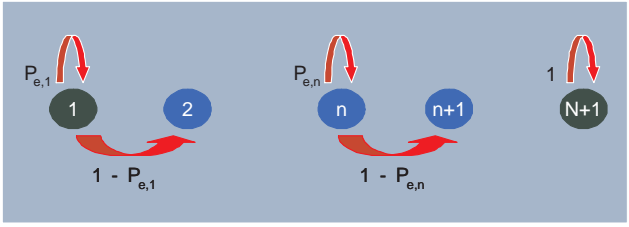}
\end{center}
\caption{Markov chain model to characterize communication over the ARQ-supported linear multi-hop network.}
\label{linear_arq}
\end{figure}

\section{End-to-End Latency of Communication over ARQ-Supported Multi-hop Network} 

Using the reliability bounds developed for the linear multi-hop network in Section II, we can provide further insights toward the rate-reliability-delay tradeoff, by introducing an automatic repeat request (ARQ) mechanism applicable over any given hop upon decoding failures due to transmission errors. In this setting, an ARQ protocol is considered, where, upon detection of codeword error (e.g., practical systems typically use a cyclic redundancy check (CRC) code), the erroneous codeword is discarded by the receiver and the retransmission of the codeword is requested from the transmitter. At any given hop, the retransmission request is repeated until the decoder detects an error-free transmission. It is assumed that the channel states $\{\theta_n\}_{n=1}^N$ do not change during retransmissions. 

The communication over the linear multi-hop network under the support of the described ARQ mechanism can be characterized using a finite-state Markov chain model \footnote{The Markov property is satisfied based on the facts that the transmit symbols $\{s_{n,q}\}$ are drawn in an independent and identically distributed (i.i.d.) fashion across $q=1,...,Q_n$ based on the probability density $p(s)$, the channel states $\{\theta_n\}_{n=1}^N$ across multiple hops are drawn in an i.i.d. fashion and that for any given state $\theta_n$, the channel over hop $n$ that maps the input symbol vector ${\bf s}_n$ to the output symbol vector ${\bf y}_n$ acts i.i.d. across $q=1,...,Q_n$ (based on the DTMC model) dictated by the conditional probability density function $p(y|s,\theta_n)$. } with a discrete-time stochastic process $Z_j,\, j=1,2,...$, which has $N+1$ states indexed by $n=1,...,N+1$ as depicted in Fig. \ref{linear_arq}, where $j$ is the code block transmission index. The transition from state $n$ to state $n+1$ represents the transmissions from terminal ${\cal T}_n$ to terminal ${\cal T}_{n+1}$ over hop $n$ and each transmission could result in a success which means that the Markov chain arrives at state $n+1$ or in a failure which means that the Markov chain remains at state $n$. The state-transition probabilities are functions of the codeword error probabilities $P_{e,n}$, which stay constant over retransmissions (since $\{\theta_n\}_{n=1}^N$ do not change). Arrival at state $N+1$ implies successful decoding of the message by the destination terminal ${\cal T}_{N+1}$, and thus this state is modeled as an absorption state; which means that the Markov chain terminates upon entering this state (i.e., no more transmissions are necessary), whereas states $1,...,N$ are transient. 

Our objective is use the described Markov chain model to compute the expected value of end-to-end latency $T$ in terms of the total required number of channel uses until successful reception of the message by the destination terminal. Toward this goal, we define the stopping time $J$ of the Markov process as $$J = \min \{j \geq 1: Z_j = N+1 \,|\, Z_1 = 1\},$$ based on which $T$ can be represented as $$T = {\Bbb E} \left[ \sum_{j=1}^{J-1} Q_{Z_j} \,|\, Z_1 = 1 \right].$$ 
This expectation can easily be computed by using the well-known first-step analysis technique based on the application of the law of total probability, exploting the Markov property of the process $Z_j$. Now, defining $T_n$ to be the expected number of channel uses until the message arrives at state $N+1$ given that the message is currently at state $n$, expressed as
$$T_n = {\Bbb E} \left[ \sum_{j=1}^{J-1} Q_{Z_j} \,|\, Z_1 = n \right],\,\,n=1,...,N+1,$$ 
we can specify the end-to-end expected latency over the multi-hop network for the block allocation $\{Q_n\}_{n=1}^N$ by the set of recursive relations (by conditioning on the outcome of the next transmission)
\begin{eqnarray}
T_{n} & = & (T_{n} + Q_n) P_{e,n} + (T_{n+1} + Q_n) (1-P_{e,n}) \nonumber\\
& = & Q_n + T_n P_{e,n} + T_{n+1}(1-P_{e,n}),\,\,\,\,\,n=1,...,N \nonumber
\end{eqnarray}
under the constraint $T_{N+1}=0$. Solving for $T$, we obtain
\begin{equation}
T(\{R_n\}_{n=1}^N,\{\theta_n\}_{n=1}^N) = \sum_{n=1}^N \frac{Q_n}{1-P_{e,n}(Q_n,R_n,\theta_n)},
\label{laten}
\end{equation}
as the expected value of end-to-end latency of multi-hop communication for the set of per-hop rates $\{R_n\}_{n=1}^N$, channel states $\{\theta_n\}_{n=1}^N$ and coding blocklengths $\{Q_n\}_{n=1}^N$.
In Section IV, we will use the upper and lower bounds on $P_{e,n}$ based on error exponents and sphere packing exponents, respectively, to bound the end-to-end latency of transmissions over the ARQ-supported linear multi-hop network given by (\ref{laten}).

\section{Additive White Gaussian Noise Channels}

Under the additive white Gaussian noise (AWGN) model, the discrete-time memoryless complex baseband input-output channel relation over the $n^{\mathrm{th}}$ hop is given by
$$
{y}_{n} =  s_n + z_{n},\,\,\,\,\,\,\,\,\,n=1,...,N,
$$
where $y_{n} \in {\Bbb C}$ is the received signal at terminal ${\cal T}_{n+1}$,
$s_n \in {\Bbb C}$ is the temporally i.i.d. zero-mean
circularly symmetric complex Gaussian scalar transmit signal from ${\cal T}_n$
satisfying the average power constraint ${\Bbb E}\left[|s_n|^2\right] = P_n$,
$z_{n} \in {\Bbb C}$ is the temporally white zero-mean circularly symmetric
complex Gaussian noise signal at ${\cal T}_{n+1}$, independent across $n$ and independent from the input signals $\{s_n\}_{n=1}^N$,
with variance $\sigma^2$. 

Unfortunately the exact computation of error exponents is highly intractable for most (including AWGN) channels mainly because of the lack of knowledge about the optimal input distribution $p(s)$. Thus, we fix the input to be i.i.d. zero-mean complex-valued Gaussian, which provides
$$
E_{0,n}(\rho,\mathsf{SNR}_n) = \rho \ln \left( 1 + \frac{\mathsf{SNR}_n}{1+\rho} \right)\,\,\,\,\,\,n=1,...,N
$$
where $\mathsf{SNR}_n$ is the average received signal to noise ratio defined as $\mathsf{SNR}_n=\frac{P_n}{\sigma^2}$ and its variation across multiple hops is dictated by the channel state $\theta_n$. To choose $p(s)$ as Gaussian is not optimal, and a distribution concentrated on a "thin spherical shell" will give better results. Nevertheless, the above expression is a convenient lower bound on $E_{0,n}$ and thus yields an upper bound to the probability of error. Maximizing the error exponent $E_{r,n}(R_n,\mathsf{SNR}_n)$ over $\rho$ according to (\ref{err1})-(\ref{err3}), we obtain the parametric equations
\begin{equation}
E_r(R_n,\mathsf{SNR}_n) = \frac{ \rho_n^2 \mathsf{SNR}_n} { (1+\rho_n)(1+\rho_n+ \mathsf{SNR}_n)},
\label{err1_g}
\end{equation}
and (\ref{err2_g})
\begin{figure*}[!ht]
\begin{equation}
R_n = \ln \left( 1 + \frac{\mathsf{SNR}_n}{1+\rho_n} \right) - \frac{\rho_n \mathsf{SNR}_n}{(1+\rho_n)(1+\rho_n+\mathsf{SNR}_n)},\,\,\,0 \leq \rho_n \leq 1
\label{err2_g}
\end{equation}
\hrule
\end{figure*}
where the latter represents the relation between the achievable rates $\{R_n\}_{n=1}^N$ and optimal choice of parameters $\left\{\rho_n\right\}_{n=1}^N$. For rates lower than those where $\rho_n = 1$, we find (\ref{err3_g}).
\begin{figure*}[!ht]
\begin{equation}
E_r(R_n,\mathsf{SNR}_n) = \ln \left( 1 + \frac{\mathsf{SNR}_n}{2} \right)-R_n,\,\,\,\,
\mathrm{for}\,\,\,\,\,R_n \leq \ln \left( 1 + \frac{\mathsf{SNR}_n}{2} \right) - \frac{\mathsf{SNR}_n}{2(2+\mathsf{SNR}_n)}
\label{err3_g}
\end{equation}
\hrule
\end{figure*}
Similarly, using (\ref{sp1})-(\ref{sp2}), we obtain the following parametric equations for the sphere packing exponent:
\begin{equation}
E_{sp}(R_n,\mathsf{SNR}_n) = \frac{ \rho_n^2 \mathsf{SNR}_n} { (1+\rho_n)(1+\rho_n+ \mathsf{SNR}_n)},
\label{sp1_g}
\end{equation}
and
\begin{equation}
R_n = \ln \left( 1 + \frac{\mathsf{SNR}_n}{1+\rho_n} \right) - \frac{\rho_n \mathsf{SNR}_n}{(1+\rho_n)(1+\rho_n+\mathsf{SNR}_n)},\,\,\,\,\forall \rho_n.
\label{sp2_g}
\end{equation}

\begin{figure}[t]  

 \centering

  \includegraphics[height=!,width=3.5in]{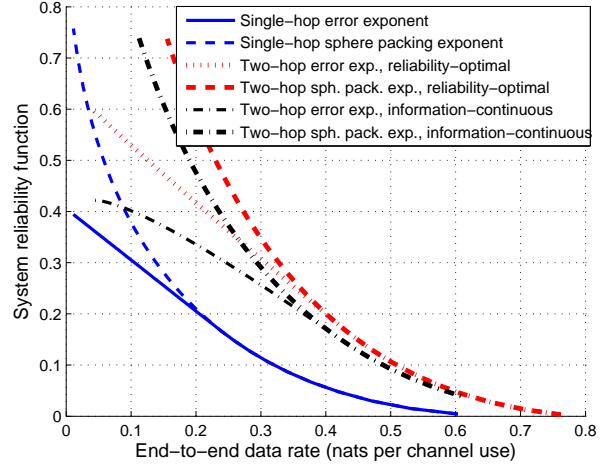}

  \caption{Lower and upper bounds on the system reliability function based on error exponents and sphere packing exponents, respectively, for single-hop and two-hop communications over an AWGN linear network.}

  \label{reliab}

\end{figure}

\begin{figure}[t]  

 \centering

  \includegraphics[height=!,width=3.5in]{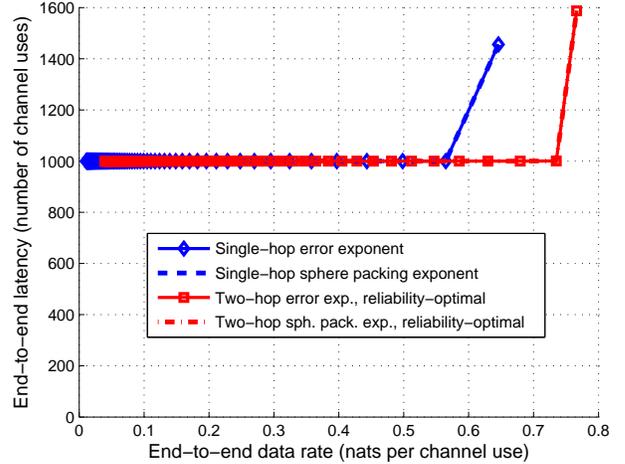}

  \caption{Upper and lower bounds on the expected value of end-to-end latency based on error exponents and sphere packing exponents, respectively, for single-hop and two-hop communications over an AWGN linear network.}

  \label{latency}

\end{figure}

In Fig. \ref{reliab}, we plot the random coding error exponent lower bound in (\ref{err_exp}) and sphere packing exponent upper bound in (\ref{sp_exp}) on the system reliability function $E(\{R_n\}_{n=1}^N)$ defined in (\ref{sys_rel}) for an AWGN linear multi-hop network as a function of end-to-end data rate. We utilize from the results in (\ref{err1_g})-(\ref{err3_g}) for the error exponent computation and from (\ref{sp1_g})-(\ref{sp2_g}) for the sphere packing exponent computation. We constrain the end-to-end delay by setting $Q$ as $Q=1000$ channel uses. As shown in \cite{Laneman05}-\cite{Oyman06b}, in relay-assisted multi-hop wireless communication networks, the signal-to-noise ratio over each link increases due to lower interterminal distances and reduced propagation path loss. We focus on the single-hop case ($N=1$) with $\mathsf{SNR}_1=0~\mathrm{dB}$ and two-hop case ($N=2$) with $\mathsf{SNR}_1=9~\mathrm{dB}$ and $\mathsf{SNR}_2=6~\mathrm{dB}$, which could represent typical SNR values for a communication link between a source-destination pair terminals where, in the single-hop case no relay terminal is present, and in the two-hop case the relay terminal placed at the midpoint between the source and destination terminals and the path loss exponent is in the range of $2$-$4$ over multiple hops. In the two-hop case, we choose the blocklengths $Q_1$ and $Q_2$ considering two different methods discussed in Section II, (i) reliability-optimal (i.e., error-minimal) solution based on (\ref{opt_blk}) and (ii) information-continous solution with $Q_n = \lfloor \frac{\ln M}{R_n} \rfloor, \,n=1,2$, where $M$ is given by (\ref{num_words}). The end-to-end data rate (in nats per channel use) is calculated as $$R = \frac{1}{Q} \min_{n=1,...,N} \{Q_nR_n\},$$ for multi-hop transmissions. We clearly see from Fig. \ref{reliab} that the path loss reduction achieved by multi-hop communication also improves the reliability function as the upper and lower bounds for the two-hop case are significantly better than those for the single-hop case. This is remarkable, especially considering the fact that the end-to-end delay level is the same for both systems under the constraint $\sum_{n=1}^N Q_n = Q$. In addition, we observe that under a fixed end-to-end delay constraint, information-continuous allocation of blocks yields a reliability that is close to that yielded by error-minimal block allocation, especially for the range of rates close to capacity. It should be mentioned that, consistent with the capacity results of \cite{Laneman05}-\cite{Oyman06b}, the superior performance of multi-hop communication is at the low SNR ranges only, and for high SNR values, single-hop communication yields a better rate-reliability-delay tradeoff.

Under the same set of assumptions, we use (\ref{laten}) to plot in Fig. \ref{latency} the expected value of end-to-end latency for single-hop and two-hop communication schemes as a function of end-to-end data rate for reliability-optimal block allocation with the support of the ARQ protocol described in Section III. We observe that the upper and lower bounds on the expected value of the end-to-end latency obtained by error exponents and sphere packing exponents, respectively, overlap with each other in full agreement, indicating the accuracy of our latency characterization. From this result, it is clear that the two-hop scheme outperforms the single-hop scheme in terms of the end-to-end latency performance in the presence of ARQ-based retransmission mechanisms, which is another evidence for the enhancement of the rate-reliability-delay tradeoff through multi-hop communication in the low SNR regime. In contrast, it should be noted that the numerical results obtained through the evaluation of (\ref{laten}) in the high SNR regime indicate that single-hop communication is preferable over multi-hop communication and hence, once again, the observed trends are consistent with the insights obtained from the capacity analysis of \cite{Laneman05}-\cite{Oyman06b}.

\section{Conclusions}

We characterized the end-to-end rate-reliability-delay tradeoff over a linear multi-hop network composed of multi-state discrete memoryless channels over each hop, with orthogonal time-sharing across hops under a half-duplex relaying protocol. Based on this general framework, we provided numerical results on the performance comparison between multi-hop and single-hop communication focusing on AWGN channels. Our analysis has led to the following conclusions:
\begin{itemize}
\item Multi-hop communication yields better end-to-end reliability over single-hop communication in the low SNR regime for a given set of achievable per-hop data rates under an end-to-end delay constraint.
\item Information-continuous block allocation yields an error performance close to that achieved by reliability-optimal block allocation under the imposed end-to-end delay constraint, especially for the set of per-hop rates close to capacity.  
\item Multi-hop communication outperforms single-hop communication in terms of the end-to-end latency performance in the low SNR regime in the presence of ARQ-based retransmission mechanisms.
\end{itemize}

\begin{footnotesize}
\renewcommand{\baselinestretch}{0.2}
\bibliographystyle{IEEE}

\end{footnotesize}

\end{document}